\documentclass[floatfix,nofootinbib]{revtex4}

\usepackage{amssymb}
\usepackage{amsmath}
\usepackage[dvips]{graphicx}

\newcommand{\be}{\begin{equation}}
\newcommand{\ee}{\end{equation}}
\newcommand{\bse}{\begin{subequations}}
\newcommand{\ese}{\end{subequations}}

\newcommand{\eq}[1]{eq.~(\ref{#1})}
\newcommand{\eqs}[1]{eqs.~(\ref{#1})}

\DeclareMathOperator{\tr}{tr}
\DeclareMathOperator{\diag}{diag}

\DeclareMathOperator{\spa}{span}

\DeclareMathOperator{\sign}{sign}

\newcommand{\ZZ}{\mathbb{Z}}
\newcommand{\RR}{\mathbb{R}}
\newcommand{\CC}{\mathbb{C}}

\newcommand{\lieg}{\mathcal{G}}
\newcommand{\cte}{c}

\newcommand{\At}{\tilde{A}}

\begin{document}

\title{Nonsingular solutions of Hitchin's equations for noncompact gauge groups}

\author{Ricardo A. Mosna}
\email{mosna@ime.unicamp.br}
\author{Marcos Jardim}
\email{jardim@ime.unicamp.br}
\affiliation{
Instituto de Matem\'atica, Estat\'\i stica e Computa\c{c}\~ao Cient\'\i fica,
Universidade Estadual de Campinas, C.P. 6065, 13083-859,
Campinas, SP, Brazil.}


\begin{abstract}
We consider a general ansatz for solving the 2-dimensional Hitchin's equations,
which arise as dimensional reduction of the 4-dimensional self-dual
Yang-Mills equations, with remarkable integrability properties. We
focus on the case when the gauge group $G$ is given by a real form
of $SL(2,\CC)$. For $G=SO(2,1)$,
the resulting field equations are shown to reduce to either the Liouville, elliptic
sinh-Gordon or elliptic sine-Gordon equations. As opposed to the compact case,
given by $G=SU(2)$, the field equations associated with the noncompact group $SO(2,1)$ are shown
to have smooth real solutions with nonsingular action densities,
which are furthermore localized in some sense.
We conclude by discussing some particular solutions, defined on $\RR^2$,
$S^2$ and $T^2$, that come out of this ansatz.
\end{abstract}

\maketitle

\section{Introduction}

Yang-Mills theory has been a rich source of profound mathematical results in the past
three decades. It has found applications in a wide variety of research areas, such as
differential topology and algebraic geometry \cite{DK}, representation theory \cite{Nak}
and in the theory of integrable systems \cite{H2,MW}.

Usually, one considers the Yang-Mills self-duality equations for connections taking
values on the Lie algebra of a compact real Lie group. So far, Yang-Mills instantons
for complex or noncompact real Lie groups have received little attention (see for instance \cite{R})
and, to our knowledge, no general theory has been developed and the proper physical interpretation
of such theory is yet to be understood, see \cite{gabor} for a recent result in this direction.
In this paper, we intend to provide motivation for the study of Yang-Mills theory with noncompact real
Lie groups by remarking upon some rather interesting phenomena taking place on a two-dimensional gauge theory.

More precisely, let $G$ be a real Lie group with Lie algebra $\lieg$, and let $G^{\CC}$
be its complexification. Let $A=A_1 dx + A_2 dy + A_3 du + A_4 dv$ be a $G$-connection 1-form
defined on Euclidean $\RR^4$, with Cartesian coordinates $(x,y,u,v)$. We are interested in the dimensionally
reduced case where the fields depend only on $(x,y)$. In this case, it is known that the
self-duality equations reduce to the so-called Hitchin's equations \cite{hitchin}:
\bse
\label{hit}
\begin{align}
F_{\At} &= - [\Phi,\Phi^*], \label{hit1}\\
d_{\At}\Phi &= 0,  \label{hit2}
\end{align}
\ese
where $\At=A_1 dx + A_2 dy$ defines a connection over $\RR^2$,
$F_{\At}$ is the curvature of $\At$
and $\Phi=\frac{1}{2}(A_3- iA_4)(dx+idy)$ can be thought of as a Higgs field.
The anti-automorphism $*$ acts on $\lieg\otimes\Omega^1_{\RR^2}\otimes\CC$ as
minus the identity on the first factor, $\lieg$, and as complex
conjugation on the space of complex valued 1-forms, $\Omega^1_{\RR^2}\otimes\CC$.
Equations (\ref{hit}) inherit the
conformal invariance of the 4-dimensional Yang-Mills equations and can be
considered on any Riemann surface $\Sigma$. It has been remarked that there are
no smooth real solutions of the these equations for $\Sigma=\RR^2,S^2,T^2$
(where $T^2$ is the 2-torus) and $G=SU(2)$~\cite{saclioglu, hitchin}.

In this paper, we obtain smooth real solutions of (\ref{hit}) for $\Sigma=\RR^2,S^2,T^2$
and $G=SO(2,1)$, with nonsingular bounded action densities.
It should be noted that the similar problem of finding solutions to the dimensionally reduced
self-duality equations taking values in the complexified version $su(2)\otimes\CC$ of $su(2)$
was addressed in the earlier~\cite{saclioglu}. Here, by employing  a result of Donaldson~\cite{donaldson}
which relates Hitchin pairs $(\At,\Phi)$ to flat $G^\CC$-connections on $\Sigma$,
we obtain an ansatz which turns out to be essentially the same as that of \cite{saclioglu}.
We show that this ansatz reduces the Hitchin's equations to some classical
integrable equations, namely the Liouville, the elliptic sinh-Gordon and the elliptic sine-Gordon equations.
The first case (Liouville) was addressed, from a different perspective, already in \cite{saclioglu},
where the author also showed that the strictly compact case
leads to the sinh-Gordon equation (but in a different variant from that obtained here
for $SO(2,1)$; see section \ref{sec:integration}).
Moreover, we note that the strong link between Yang-Mills instantons
and classical integrable equations is well known (see the book \cite{MW}),
but that here it arises within the context of noncompact real Lie groups.
The $U(n)$ Hitchin's equations also lead to a celebrated completely integrable system,
the Hitchin system~\cite{H2}.
It is remarkable that this link between Hitchin's equations and integrable systems also
arises when noncompact real Lie groups are considered.

\section{Algebraic preliminaries\label{alg}}

Let $\{\tau_1,\tau_2,\tau_3\}$ span the Lie algebra of a {\em real form} $G$ of the complex group
$SL(2,\CC)$, with
\bse
\label{tau.cr}
\begin{align}
[\tau_2,\tau_3] &= (-1)^{n_1} \tau_1, \\
[\tau_3,\tau_1] &= (-1)^{n_2} \tau_2, \\
[\tau_1,\tau_2] &= \tau_3.
\end{align}
\ese
Note that, depending on the choice of $n_1, n_2\in \{0,1\}$, $\lieg$ is isomorphic
(as a real Lie algebra) to either $su(2)$ or $so(2,1)$. In this way, the elements
$\tau_k$ are generators of $G=SU(2)$ or $SO(2,1)$ inside $SL(2,\CC)$. Let
$E_k=\frac{1}{2i}\sigma_k$ be the usual generators of $su(2)$, where $\sigma_k$
are the Pauli matrices, so that
\begin{align*}
[E_2,E_3] &= E_1, \\
[E_3,E_1] &= E_2, \\
[E_1,E_2] &= E_3.
\end{align*}
A representation of $\tau_1,\tau_2,\tau_3$ satisfying \eqs{tau.cr} is then given by
\bse
\label{tau.def}
\begin{align}
\tau_1 &= i^{n_2} E_1, \\
\tau_2 &= i^{n_1} E_2, \\
\tau_3 &= i^{n_1+n_2} E_3.
\end{align}
\ese
We note that the above generators of $\lieg$ are orthogonal with respect to its
corresponding Killing form, so that
\[
\tr(\tau_i\tau_j)=-\frac{1}{2}\,g_{ij},
\]
where the entries of the diagonal matrix $(g_{ij})$ are given by $\pm1$, with
the signs fixed as follows.
\[
\begin{tabular}{|c|c|c|c|c|}
  \hline
  $\left( n_1,n_2 \right)$ & $(0,0)$ & $(1,0)$ & $(0,1)$  & $(1,1)$  \\
  \hline $\diag(g_{ij})$   & $+++$   & $+--$   & $-+-$    & $--+$    \\
  \hline
\end{tabular}
\]

\section{Ansatz\label{ans}}

As noticed already by Hitchin in \cite{hitchin}, whenever the pair $(\At,\Phi)$ is
a solution to \eqs{hit}, the associated $\lieg_\CC$-valued 1-form $B=\At + \Phi+\Phi^*$
defines a {\em flat} $G^\CC$-connection on $\Sigma$. As shown by Donaldson, the converse
also holds when $G=SO(3)$ \cite{donaldson}. These facts have been used as
the starting point of recent work  \cite{gabor} on the physical interpretation of
the $SO(3)$ Hitchin's equations as a classical $2 + 1$ dimensional vacuum general
relativity theory on $\Sigma\times\RR$, where $\Sigma$ is a Riemann surface of genus
$g>1$.

Motivated by these general observations, and in order to obtain a reasonable {\em ansatz} for
solving \eqs{hit}, we demand that whenever $(\At,\Phi)$ is a Hitchin pair  associated with a
{\em real form} $G$ of $SL(2,\CC)$, the corresponding $sl(2,\CC)$-valued 1-form
$B=\At + \Phi+\Phi^*$ be a {\em real} flat connection, i.e., a flat connection associated
with a (possibly different) {\em real} form $G'$ of $SL(2,\CC)$.

Recall that $\At$ and $\Phi$ are defined by
$\At=A_1 dx + A_2 dy$ and $\Phi=\frac{1}{2}(A_3- iA_4)dz$, where $A_k$ takes
values in $\lieg$ and $dz=dx+idy$. In this way, $\At$ and $\Phi+\Phi^*$ take values in
$\lieg$ and $i\lieg$, respectively. Assuming that $\Phi$ is not identically zero,
it follows that the image of $\At$ and $\Phi$ will
close a Lie algebra $\lieg'$, associated with a {\em real} form $G'$, only if $A_1,A_2$ take
values in a one-dimensional subalgebra $\mathcal{A}$ of $\lieg$ (generated by $\tau_1$, say),
while $A_3,A_4$ take values in its orthogonal complement $\mathcal{A}^\perp$
(generated by $\{\tau_2,\tau_3\}$, say) inside $\lieg$. In this way,
the image of $B=\At + \Phi+\Phi^*$ will be contained in the {\em real} Lie algebra
(i.e., with {\em real} structure constants) $\lieg'=\spa_{\RR}\{\tau_1,i\tau_2,i\tau_3\}$.

This leads to the following {\em ansatz} for $A$:
\be
A=\left( f_1 dx + f_2 \, dy \right) \, \tau_1 +
\left( g_2\tau_2 + g_3\tau_3 \right) du +
\left( h_2\tau_2 + h_3\tau_3 \right) dv,
\ee
where $f_k,g_k,h_k$ are functions of $(x,y)$ and $\tau_1,\tau_2,\tau_3$ are as above.
An appropriate gauge transformation $A\to q^{-1} A q + q^{-1}dq$, with $q(x,y)=e^{\chi(x,y)\tau_1}$,
can be used to gauge away $g_3$. The particular case of $h_2=0$ then leads to
\be\label{ansatz}
A=\left( f_1 dx + f_2 \, dy \right) \, \tau_1 + g \, du \, \tau_2 + h \, dv \, \tau_3.
\ee
As noted in the introduction, this ansatz was already considered in \cite{saclioglu} and,
for the sake of simplicity, we also restrict ourselves to it in what follows.
However, we note that a nonzero $h_3$ is allowed by the above discussion and this may
eventually lead to interesting and more general results.

Let $F=dA+A\wedge A$ be the curvature 2-form associated with $A$.
A straightforward calculation shows that $F=\frac{1}{2}F_{\mu\nu}dx^\mu\wedge dx^\nu$ is given by
\bse
\label{Fmunu}
\begin{align}
F_{12} &= \left( \partial_x f_2 - \partial_y f_1 \right) \tau_1, \\
F_{13} &= \partial_x g \; \tau_2 +           f_1 g \; \tau_3, \\
F_{14} &= \partial_x h \; \tau_3 -(-1)^{n_2} f_1 h \; \tau_2, \\
F_{23} &= \partial_y g \; \tau_2 +           f_2 g \; \tau_3, \\
F_{24} &= \partial_y h \; \tau_3 -(-1)^{n_2} f_2 h \; \tau_2, \\
F_{34} &= (-1)^{n_1} gh \; \tau_1,
\end{align}
\ese

and the Hitchin's equations (\ref{hit}) result in
\bse
\label{a}
\begin{align}
\partial_x g                    &=  (-1)^{n_2} f_2 h,  \label{a1} \\
\partial_y g                    &= -(-1)^{n_2} f_1 h,  \label{a2} \\
\partial_x h                    &=             f_2 g,  \label{a3} \\
\partial_y h                    &= -           f_1 g,  \label{a4} \\
\partial_x f_2 - \partial_y f_1 &=  (-1)^{n_1} gh.     \label{a5}
\end{align}
\ese

\subsection{Field equations and action functional}

It follows from \eqs{a1} and (\ref{a3}) that
$\partial_x \left[ (-1)^{n_2} g^2 -  h^2  \right]=0$.
Similarly, it follows from \eqs{a2} and (\ref{a4}) that
$\partial_y \left[ (-1)^{n_2} g^2 -  h^2  \right]=0$.
Therefore, the quantity
\be
\label{kappa}
\kappa^2 = g^2 -  (-1)^{n_2} h^2
\ee
is conserved.
Taking derivatives of \eqs{a1} and(\ref{a2}), and using the remaining
equations in order to eliminate $f_1,f_2,h$ in terms of $g$, a straightforward
calculation shows that $g$ should satisfy the following nonlinear PDE:
\be
\label{EDg}
\nabla^2 g = \frac{g}{g^2-\kappa^2} (\nabla g)^2 + (-1)^{n_1} g (g^2-\kappa^2),
\ee
where $\nabla^2=\partial_x^2+\partial_y^2$ is the Euclidean Laplacian.

The action functional
\[
S=-\frac{1}{8\pi^2}\int \tr(F\wedge\star F)
\]
has a simple expression in the context of {\em ansatz} (\ref{ansatz}).
In this case,
\be
\label{action.raw}
S=\frac{1}{8\pi^2} \int \left[
(-1)^{n_2}(gh)^2 + (-1)^{n_1+n_2}(f_1^2+f_2^2)[g^2+(-1)^{n_2}h^2]
\right] dx\wedge dy\wedge du\wedge dv,
\ee
so that
\be
S=\frac{1}{8\pi^2} \int \left[
g^2(g^2-\kappa^2)+(-1)^{n_1} (\nabla g)^2 \frac{2g^2-\kappa^2}{g^2-\kappa^2}
\right] dx\wedge dy\wedge du\wedge dv,
\ee
where we made use of \eqs{a} to eliminate $f_1,f_2,h$ in terms of $g$. Substituting
the field equations (\ref{EDg}) then leads to
\be
S=\frac{(-1)^{n_1}}{16\pi^2} \int \nabla^2 g^2 \; dx\wedge dy\wedge du\wedge dv.
\label{action.a}
\ee
Since the fields under consideration depend only on the coordinates $(x,y)$,
it is useful to define the reduced action
\be
S'=\frac{(-1)^{n_1}}{16\pi^2} \int \nabla^2 g^2 \; dx\wedge dy,
\label{redaction1}
\ee
so that $S=\iint S' dudv$ (note that $S$ will always diverge unless $S'=0$),
and its associated (2-dimensional) density
\be\label{acdens}
\sigma=\frac{(-1)^{n_1}}{16\pi^2} \nabla^2 g^2.
\ee

If $g$ is regular on the plane, except (possibly) at the origin, then it
follows from the divergence theorem that
\be
\label{redaction2}
S'=\frac{(-1)^{n_1}}{16\pi^2}\left[
\lim_{r\to\infty}  \oint_{\mathcal{C}_r} \frac{\partial g^2}{\partial r} r d\theta -
\lim_{r\to0}       \oint_{\mathcal{C}_r} \frac{\partial g^2}{\partial r} r d\theta
\right].
\ee
In particular, when $g$ is a radial function, we get
\be
S'=\frac{(-1)^{n_1}}{8\pi} \left[r\frac{d g^2}{dr} \right]_0^\infty
\qquad (\text{when } g=g(r)).
\label{redaction3}
\ee

\section{Integration\label{sec:integration}}

Due to the functional form of the conserved quantity $\kappa$ in \eq{kappa}, it is useful to
consider two separate cases in order to integrate the equations of motion.

\subparagraph*{Case (i): $(-1)^{n_2}=+1$}
In this case, \eq{kappa} reduces to
\be
g^2 -  h^2 = \kappa^2.
\label{kappa.i}
\ee
We show in the following that, for $\kappa=0$, the dynamics of the system is
governed by the Liouville equation, while $\kappa\ne0$ is associated with
the sinh-Gordon equation.

We first consider the case given by $\kappa=0$. Then, $h=\pm g$ and
we may assume, without loss of generality, that $h=g$ (the other case
can be mapped into this one by redefining $f_1$ and $f_2$ to $-f_1$ and
$-f_2$ in \eqs{a}). It follows from~\eqs{a} that
\bse
\label{f.i}
\begin{align}
f_1 &=-\partial_y \ln(g), \label{f.1.a}\\
f_2 &= \partial_x \ln(g), \label{f.1.b}\\
\partial_x f_2 - \partial_y f_1 &=  (-1)^{n_1} g^2. \label{f.1.c}
\end{align}
\ese
Substitution of eqs.~(\ref{f.1.a})-(\ref{f.1.b}) into \eq{f.1.c} then yields
\[
\nabla^2(\ln g) -(-1)^{n_1} g^2 =0.
\]
Thus $\lambda=g^2$ satisfies the {\em Liouville equation}
\[
\nabla^2(\ln \lambda) \pm 2\lambda =0.
\]

We now consider the case when $\kappa\ne0$ in \eq{kappa.i}. Since
$g^2 - h^2 = \kappa^2$,
it is natural to write
\begin{align}
g &=\kappa \cosh(\alpha), \\
h &=\kappa \sinh(\alpha),
\end{align}
for some yet unknown function $\alpha=\alpha(x,y)$. Substitution into \eqs{a} yields
\bse
\begin{align}
f_1 &=- \partial_y \alpha, \\
f_2 &=  \partial_x \alpha, \\
\partial_x f_2 - \partial_y f_1 &=  (-1)^{n_1} \frac{\kappa^2}{2}\sinh(2\alpha).
\end{align}
\ese
Thus $\alpha$ satisfies, in this case, the {\em (elliptic) sinh-Gordon equation}
\[
\nabla^2 \alpha -(-1)^{n_1}\frac{\kappa^2}{2}\sinh(2\alpha)=0.
\]
The reduced action (\ref{redaction1}) is given, in terms of $\alpha$, by
\be
S'=\frac{(-1)^{n_1}\kappa^2}{16\pi^2} \int \nabla^2 \cosh^2(\alpha) \; dx\wedge dy.
\ee
When $\alpha$ is a radial function, it follows from \eq{redaction3} that
\be
\label{S.sinh}
S'=\frac{(-1)^{n_1}\kappa^2}{8\pi} \left[r \sinh(2\alpha)\frac{d \alpha}{dr} \right]_0^\infty
\qquad (\text{when } \alpha=\alpha(r)).
\ee

\subparagraph*{Case (ii): $(-1)^{n_2}=-1$}

In this case, \eq{kappa} reduces to
\be
g^2 +  h^2 = \kappa^2.
\label{kappa.ii}
\ee
This is nontrivial only if $\kappa\ne0$. It is then natural to write
\begin{align}
g &=\kappa \cos(\alpha), \\
h &=\kappa \sin(\alpha).
\end{align}
Substitution in  \eqs{a} yields
\bse
\begin{align}
f_1 &= - \partial_y \alpha, \\
f_2 &=   \partial_x \alpha, \\
\partial_x f_2 - \partial_y f_1 &=  (-1)^{n_1} \frac{\kappa^2}{2}\sin(2\alpha).
\end{align}
\ese
Thus
\[
\nabla^2 \alpha +(-1)^{n_1}\frac{\kappa^2}{2}\sin(2\alpha)=0,
\]
which is the {\em (elliptic) sine-Gordon equation}. The reduced action (\ref{redaction1})
is now given, in terms of $\alpha$, by
\[
S'=\frac{(-1)^{n_1}\kappa^2}{16\pi^2} \int \nabla^2 \cos^2(\alpha) \; dx\wedge dy.
\]
When $\alpha$ is a radial function, it follows from \eq{redaction3} that
\[
S'=\frac{(-1)^{n_1}\kappa^2}{8\pi} \left[r \sin(2\alpha)\frac{d \alpha}{dr} \right]_0^\infty
\qquad (\text{when } \alpha=\alpha(r)).
\]

\subsection{Example 0}
Let $(n_1,n_2)=(0,0)$ in \eqs{tau.cr}. Then $\{\tau_k\}$ spans the Lie algebra $su(2)$,
generating in this way the compact group $G=SU(2)$. The Hitchin's equations for this
case follows from the analysis of case (i) above, so that
\bse
\label{ex.su2}
\begin{align}
\kappa  =0 \quad \Rightarrow & \quad \nabla^2 \ln g -g^2=0,   \label{ex.su2.a} \\
\kappa\ne0 \quad \Rightarrow & \quad \nabla^2 \alpha - \frac{\kappa^2}{2}\sinh(2\alpha)=0.  \label{ex.su2.b}
\end{align}
\ese
Note that, in this case, the integrand of \eq{action.raw} is positive definite. Thus,
we have $S'=0$ (and consequently a finite $S$) only if $A=0$.

\subsection{Example 1}
Let $(n_1,n_2)=(1,0)$ in \eqs{tau.cr}. Then $\{\tau_k\}$ spans the Lie algebra $so(2,1)$,
generating in this case the noncompact group $G=SO(2,1)$. Note that,
in this case, $\tau_1$ generates a compact Abelian subgroup of $G=SO(2,1)$.
The Hitchin's equations for this case also follows from the analysis of case (i) above,
now with $(-1)^{n_1}=-1$, and yield
\bse
\label{ex.sinh}
\begin{align}
\kappa  =0 \quad \Rightarrow & \quad \nabla^2 \ln g + g^2=0,  \label{ex.sinh.a} \\
\kappa\ne0 \quad \Rightarrow & \quad \nabla^2 \alpha + \frac{\kappa^2}{2}\sinh(2\alpha)=0. \label{ex.sinh.b}
\end{align}
\ese
In this case one may have, in principle, nontrivial solutions of \eq{hit} with finite action
(notice that the Killing form of $SO(2,1)$ is not positive definite, unlike that of $SU(2)$).
Note the all-important difference in the $\pm$ sign appearing in
\eqs{ex.su2} and \eqs{ex.sinh}. As mentioned in the introduction,
although the earlier work \cite{saclioglu} considers the dimensionally reduced
Yang-Mills equations with complexified Lie algebra, the sinh-Gordon equation appears
there only in the variant of example 0 which, as discussed above, necessarily
yields singular real solutions in $S^2$ or $\RR^2$. The same variant of the
sinh-Gordon (i.e., \eq{ex.su2.b} of example 0) also appears in the context of Matrix
String Theory \cite{bonora}.

\subsection{Example 2}
Let $(n_1,n_2)=(0,1)$ in \eqs{tau.cr}. Then $\{\tau_k\}$ generates, once again, the Lie algebra $so(2,1)$.
Now $\tau_1$ generates a noncompact Abelian subgroup of $G=SO(2,1)$.
The Hitchin's equations for this case now follows from the case (ii) above,
and yield the elliptic sine-Gordon equation
\be
\label{ex.sine}
\nabla^2 \alpha + \frac{\kappa^2}{2}\sin(2\alpha)=0.
\ee

\section{Smooth solutions in $S^2$ for $G=SO(2,1)$}

This section contains a brief discussion on solutions of the Hitchin's equations
defined on $S^2$. Its content is largely not new, since most of the material appearing
here was considered, from a different perspective, in \cite{saclioglu}
(as mentioned in the introduction). We nonetheless discuss this case here in order to contrast it to
the results of the next section, where we consider solutions defined on $\RR^2$ which
are not extendable to $S^2$.

As discussed in section~\ref{sec:integration}, when $\kappa=0$ the
Hitchin's equations reduce to the Liouville equation.
It follows from \eqs{ex.su2.a} and (\ref{ex.sinh.a}) that, in this case,
\be
\nabla^2(\ln \lambda) \pm 2\lambda =0,
\label{le}
\ee
where $\lambda=g^2$. The top sign corresponds to example 1, where $\tau_1$ generates
an Abelian subalgebra of the noncompact group $G=SO(2,1)$, while the bottom sign
corresponds to the compact case $G=SU(2)$ of example 0. In terms of $z=x+iy$, \eq{le} reads
\[
\partial_z\partial_{\bar{z}}(\ln \lambda)\pm\frac{1}{2}\lambda=0,
\]
and its general solution (found by Liouville already in 1853), is
given by \cite{liouville}
\begin{equation}
\lambda(z,\bar{z})=4\frac{\xi'(z) \eta'(\bar{z})}{[1\pm \xi(z)\eta(\bar{z})]^2},
\end{equation}
where $\xi(z)$ and $\eta(\bar{z})$ are arbitrary analytic and anti-analytic functions,
respectively. Real solutions can be obtained by setting $\eta(\bar{z})=\overline{\xi(z)}$
(but see the appendix of \cite{swe}). The choice $\xi(z)=z^\nu$ then
leads to axisymmetric solutions, where the fields depend on the radial
coordinate $r$ only. In terms of polar coordinates $z=r e^{i\theta}$, and recalling
that $\lambda=g^2$, one gets
\be
g^2(r)=4\nu^2\frac{r^{2\nu-2}}{(1\pm r^{2\nu})^2}.
\label{g^2.a}
\ee
It follows from \eqs{f.i} that, in this case,
\bse
\label{sol.a}
\begin{align}
\At     &= \left( \nu-1 \mp 2\nu \frac{r^{2\nu}}{1\pm r^{2\nu}} \right) \tau_1 \, d\theta, \label{sol.a.1}\\
\Phi &= |\nu| \frac{r^{\nu-1}}{|1\pm r^{2\nu}|} \left( \tau_2 -i\tau_3 \right)\, dz. \label{sol.a.2}
\end{align}
\ese
It is interesting to note that, for the particular case of $\nu=1$,
$\At = -2 \frac{r^{2}}{1+ r^{2}}\,d\theta$ corresponds
to an Abelian magnetic monopole with charge $-2$.

In the remainder of this section we restrict ourselves to the noncompact case, which is given by the
top sign in the equations above.
We note that $\At$ from \eq{sol.a.1} is in general singular both at $r=0$ and $r\to\infty$.
By making gauge transformations on $A$, $A\to q^{-1}Aq+q^{-1}dq$, with $q=e^{N\theta\tau_1}$,
and choosing, respectively, $N=-\nu+1$ and $N=\nu+1$, we obtain
\bse
\label{sol.reg0}
\begin{align}
\At'     &= i\nu \frac{(z\bar{z})^\nu}{1 + (z\bar{z})^\nu}
             \left(  \frac{dz}{z}-\frac{d\bar{z}}{\bar{z}}  \right) \tau_1, \label{sol.reg0.1}\\
\Phi' &= |\nu| \frac{z^{\nu-1}}{1 + (z\bar{z})^{\nu}} dz
            \left( \tau_2 -i\tau_3 \right), \label{sol.reg0.2}
\end{align}
\ese
which is regular at the origin, and
\bse
\label{sol.reginf}
\begin{align}
\At''  &= -i\nu \frac{1}{1+(z\bar{z})^\nu}
          \left(  \frac{dz}{z}-\frac{d\bar{z}}{\bar{z}}  \right) \tau_1, \label{sol.reginf.1}\\
\Phi'' &= |\nu|\frac{1}{z\bar{z}}\frac{\bar{z}^{\nu+1}}{1 + (z\bar{z})^\nu} \, dz
          \left( \tau_2 -i\tau_3 \right), \label{sol.reginf.2}
\end{align}
\ese
which is regular at the infinity. After the obvious transformation $z\mapsto1/z$ in the
second set of equations, \eqs{sol.reg0} and (\ref{sol.reginf}) patch
together to define a Hitchin pair on the Riemann sphere $S^2$. We note that
the gauge transformation relating these two sets of fields is given by
$e^{2\theta\nu}$. Although the argument of $e^{2\theta\nu}$ allows semi-integer values for $\nu$,
terms like $z^{\nu-1}$ that appear in the Higgs fields above are only compatible
with integer values of $\nu$. Therefore, the above expressions define a solution of the
Hitchin's equations on $S^2$ whenever $\nu$ assumes integer values.
The reduced action $S'$ (\eq{redaction3}) is zero in this case.
In contrast, a straightforward calculation shows that $S'$ always diverges
in the compact case (given by \eqs{sol.a} with the bottom sign).

Let $F_{\At} = d\At + \At\wedge\At$ be the field strength associated with $\At$.
Both $\At$ and $F_{\At}$ can be though of as effectively Abelian fields (taking values in the
$u(1)$ subalgebra of $\lieg$ generated by $\tau_1$). It follows from \eqs{sol.reg0.1} that
\[
F_{\At} = - 4\nu^2\frac{r^{2\nu-2}}{(1 + r^{2\nu})^2} \, \tau_1 \, dx\wedge dy.
\]
The magnetic field $F_{\At}$ gives rise to a flux $\varphi = \int F_{\At}= -4\pi\nu$,
and therefore the flux $\varphi$ is quantized in units of $4\pi$.

Alternatively, one may also choose $\xi(z)$ to be a degree $n$ polynomial of the form
$(z-z_1)\cdots(z-z_n)$, $z_1,\dots,z_n\in\CC$, leading to multi-centered solutions which
were also considered in \cite{saclioglu}. We therefore conclude that the moduli space of
smooth solutions the $SO(2,1)$ Hitchin's equations on $S^2$ possesses an infinite number
of connected components which are parametrized by a positive integer $n\ge1$ (the flux),
and that each connected component has dimension at least $n$.

\section{Axisymmetric solutions in $\RR^2$}

In this section, we discuss smooth solutions of the $SO(2,1)$ Hitchin's equation on $\RR^2$
which, unlike those in the previous section, are not extendable to~$S^2$.

\subsection{Sinh-Gordon equation}
\label{subsection:sinh}

We now consider the case (i) of section~\ref{sec:integration}, with $\kappa\ne0$.
For $\alpha=\alpha(r)$, \eqs{ex.su2.b} and (\ref{ex.sinh.b}) lead to
\be
\frac{d^2 \alpha}{dr^2} + \frac{1}{r} \frac{d\alpha}{dr} \pm \frac{\kappa^2}{2}\sinh(2\alpha)=0,
\label{EDsinh}
\ee
where the convention for the top/bottom sign is the same as in the previous section.

It is interesting to note that the change of variables $U(r)=e^{\alpha(r)}$
turns the nonlinear ODE~(\ref{EDsinh}) into
\be
\frac{d^2 U}{dr^2}-\frac{1}{U}\left(\frac{dU}{dr}\right)^2+\frac{1}{r}\frac{dU}{dr}=
\mp \frac{\kappa^2}{4}\left(U^3-\frac{1}{U}\right),
\label{edforalpha.sinh.p}
\ee
which is an instance of the Painlev\'e equation of the third kind, with
$\alpha=\beta=0$ and $\gamma=-\delta=\mp\kappa^2/4$ (in the classification of \cite{ince}).
The connection to Painlev\'e III in the case $\alpha=\beta=0$ and $\gamma=-\delta=\kappa^2/4$
(bottom sign of \eq{edforalpha.sinh.p}) was already noted in \cite{saclioglu}.
It is well known that the Painlev\'e equations do not admit, in general, solutions
in ``closed form''. This motivates us to first obtain qualitative and then numerical
information from \eq{EDsinh}.

\subsubsection{Asymptotic behavior}

Let us first analyze the possible asymptotic behaviors of $\alpha(r)$ as $r\to0$.
We first note that, if $\alpha(r)\to\pm\infty$ as $r\to0$,
then the asymptotic form of
\eq{EDsinh} implies that $\alpha(r)$ goes like $\mp k\ln r$, $k\in (0,1)$, for $r\ll 1$. This leads, in turn,
to a divergence of $S'$ for $r\to0$ in \eq{S.sinh}.

Let us then assume that $\alpha(r)$ is finite
when $r$ approaches $0$, so that we can expand $\alpha(r)$ in power series near the origin.
Equation~(\ref{EDsinh}) then leads to the asymptotic expression
\be
\label{alphaas}
\alpha(r) \xrightarrow[r\to0]{}
a \mp \frac{\kappa^2}{8} \sinh(2a) \, r^2 + \frac{\kappa^4}{256} \sinh(4a) \, r^4 + \mathcal{O}(r^6).
\ee
This yields a convergent expression for $S'$ near the origin.
Moreover, in order that the quantity between square brackets in \eq{S.sinh}
remain bounded for $r\gg1$, we demand that $\alpha(r)\to0$ as $r\to\infty$.
We note in passing that the constant solution $\alpha(r)=0$ is stable for \eq{EDsinh}
with the top sign ($G=SO(2,1)$), but unstable for \eq{EDsinh} with the bottom sign
($G=SU(2)$). In this way, the requirement that $\alpha(r)\to0$ as $r\to\infty$
implies that the unique solution for example~0 ($G=SU(2)$, bottom sign) is the trivial $\alpha(r)=0$.
For the case of example 1 ($G=SO(2,1)$, top sign), the requirement that $\alpha(r)\to0$ as $r\to\infty$
leads to the the asymptotic form
$\frac{d^2 \alpha}{dr^2} + \frac{1}{r} \frac{d\alpha}{dr} + \kappa^2 \alpha=0$ for \eq{EDsinh},
where $\sinh(2\alpha)$ was approximated by $2\alpha$ in the appropriate limit.
This is a well-known instance of the Bessel equation,
whose solutions have the asymptotic form
\[
\alpha(r) \approx \cte \frac{\sin(\kappa r - \theta_0)}{\sqrt{\kappa r}}, \quad r\to\infty,
\]
where $\cte$ and $\theta_0$ are real constants. It follows that
$r \sinh(2\alpha)\frac{d \alpha}{dr} \to \cte^2\sin(2\kappa r-2\theta_0)$ as $r\to \infty$.
In this way, although the quantity between square brackets in \eq{S.sinh}
is bounded as $r\to\infty$, it asymptotically oscillates (with frequency $2 \kappa$)
with constant amplitude as $r\to \infty$,
and thus $S'$ is not well defined in this context (but see below).
Another quantity of interest is the effectively Abelian field strength
$F_{\At}=F_{12} \, dx\wedge dy$ associated with $\At$. Under the above conditions,
it is easy to see that $F_{12}\to -\kappa^2 \cte \frac{\sin(\kappa r -\theta_0)}{\sqrt{\kappa r}}$
as $r\to\infty$. This slow decay rate renders the integral $\varphi = \int F_{\At}$ ill-defined,
but it still makes sense to consider the ``current'' $\mathbf{J}$,
given by the curl of the field strength (as in Maxwell equations),
which can be interpreted as the source of the ``magnetic field'' $F_{12}$
piercing $\Sigma=\RR^2$.
A straightforward calculation shows that $\mathbf{J}=J_\theta \, \mathbf{e_\theta}$, where
$\mathbf{e_\theta}$ is the azimuthal (unit) vector field in $\RR^2$, with
$J_\theta\to\kappa^3 \cte \frac{\cos(\kappa r-\theta_0)}{\sqrt{\kappa r}}$ as $r\to\infty$.
Note that, on account of \eqs{hit}, $\mathbf{J}$ can be entirely written in terms of the Higgs
field, $\Phi$, which is to be interpreted as the ultimate source of $F_{12}$ in this setting.
It is also interesting to note the all of these asymptotic oscillation frequencies
are determined by the conserved quantity $\kappa$.

The above semi-qualitative analysis can be justified by rigorous studies on the global
asymptotics of the third Painlev\'e equations. In fact, it is possible to show%
\footnote{\label{fn1}We thank an anonymous referee for bringing this fact to our attention.}
\cite{novo1,novo2,novo3,novo4,book} that
there exists a one parameter family of regular solutions of \eq{EDsinh} with the top sign,
defined for for $r\ge0$, characterized by the Cauchy data:
\[
\alpha(0)=a, \quad \alpha'(0)=0.
\]
These solutions have the asymptotic behavior given by
\be
\label{exact.sinh}
\alpha(r) = \cte \, \frac{\sin(\phi(\kappa r)-\theta_0)}{\sqrt{\kappa r}} + \mathcal{O}(r^{-1/2}),\quad r\to\infty,
\ee
where $\phi(z)=z+\frac{\cte^2}{4}\ln z$, and the parameters $\cte$ and $\theta_0$ are related
to the initial condition $a$ by
\bse
\label{exact.sinh.parameters}
\begin{align}
\cte^2   &= \frac{4}{\pi}\ln\cosh a,   \\
\theta_0 &= -\frac{\cte^2}{2}\ln2-\frac{\pi}{4}+\arg\Gamma\left(\frac{i\cte^2}{4}\right)+\frac{\pi}{2} \sign(a) \quad \pmod{2\pi}.
\end{align}
\ese
It is interesting to observe that our semi-qualitative analysis agrees with these exact
asymptotic expressions in the appropriate limit.
We also note that asymptotic expressions for the singular solutions of \eq{EDsinh}
with the bottom sign can be also found in~\cite{novo1,novo2,novo3,novo4,book}.

\subsubsection{Numerical results}

\begin{figure}[htbp]
\begin{center}
\includegraphics[width=15cm]{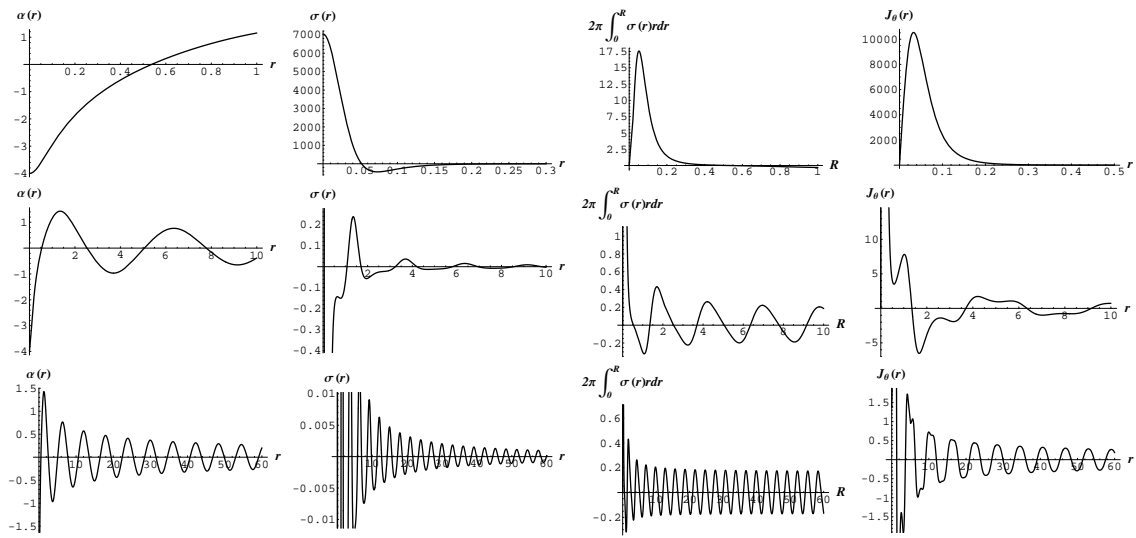}
\caption{First column: typical plots of $\alpha$ as a function of the radial coordinate $r$
for the sinh-Gordon case.
Second column: reduced action density $\sigma$ as a function of $r$.
Third column: $2\pi\int_0^R \sigma(r)\,rdr$ as a function of the upper limit $R$.
Fourth column: azimuthal component $J_\theta$ of the Abelian current $\mathbf{J}$ (see text).
}
\label{figsinh}
\end{center}
\end{figure}

\begin{figure}[htbp]
\begin{center}
\includegraphics{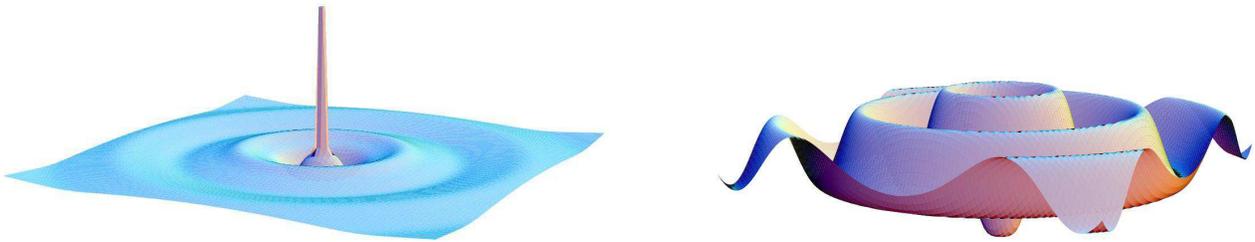}
\caption{Typical plots of $F_{12}$ for
the sinh-Gordon (left) and sine-Gordon (right) cases. The parameters are
the same as those in Figs.~\ref{figsinh} and \ref{figsine}.
}
\label{figFz3d}
\end{center}
\end{figure}

Numerical integration of (\ref{EDsinh}) starting at a given $r_0$ such that
$0<r_0\ll 1$, with $\alpha(r_0)$ and $\alpha'(r_0)$ defined by \eq{alphaas},
does support the above analysis (note that one cannot start the numerical integration
from $r=0$, since the differential equation is singular at this point).
Illustrative results for $G=SO(2,1)$ (top sign), with $r_0=10^{-3}$, $\kappa=1$ and $a=-4$,
are shown in Fig.~\ref{figsinh}. The first column exhibits
the numerical solution for $\alpha$ generated by the parameters above.
The second and third columns show,
respectively, the reduced action density $\sigma$ (\eq{acdens}) and its integral
$2\pi\int_0^R \sigma(r)\,rdr$, which reproduce the behavior expected from our previous discussion.
We note that, although the integral of the action density oscillates with
constant amplitude in the limit of large $r$ (as discussed above) it is somehow
localized near the origin. The same can be said about the ``magnetic field'' $F_{12}$,
which is plotted on the left of Fig.~\ref{figFz3d},
and its associated current $J_\theta$, which is plotted on the fourth column of Fig.~\ref{figsinh}.
This provides an alternative understanding of why the fields are not strictly
localized in this case, since their source, $J_\theta$, extends to infinity with the slow
decay profile given by $1/\sqrt{\kappa r}$ (see above discussion).

\subsection{Sine-Gordon equation}
\label{subsection:sine}

We finally consider case (ii) of section~\ref{sec:integration}.
An axisymmetric solution then satisfies, from \eq{ex.sine},
\be
\frac{d^2 \alpha}{dr^2} + \frac{1}{r} \frac{d\alpha}{dr} + \frac{\kappa^2}{2}\sin(2\alpha)=0.
\label{EDsine}
\ee
A change of variables shows that this case is also connected to the Painlev\'e equation
of the third kind. In fact, writing $V(r)=e^{i\alpha(r)}$ turns \eq{EDsine} into
\be
\frac{d^2 V}{dr^2}-\frac{1}{V}\left(\frac{dV}{dr}\right)^2+\frac{1}{r}\frac{dV}{dr}=
\frac{\kappa^2}{4}\left(V^3-\frac{1}{V}\right).
\label{edforalpha.p}
\ee

\begin{figure}[htbp]
\begin{center}
\includegraphics[width=15cm]{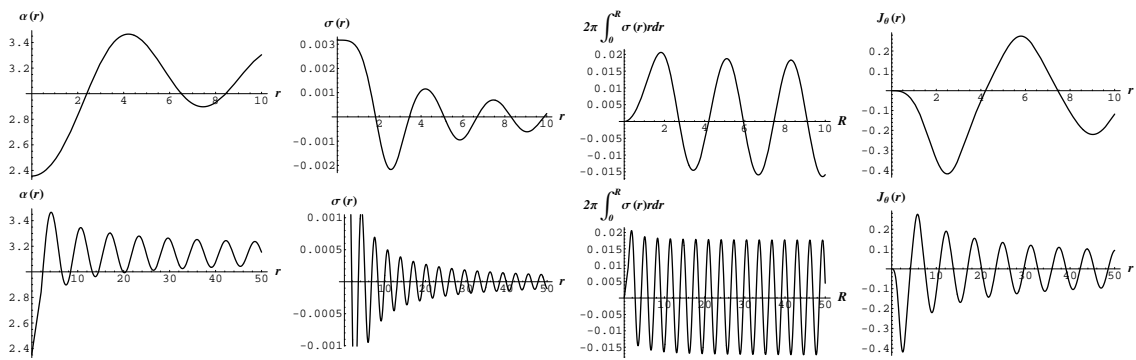}
\caption{First column: typical plots of $\alpha$ as a function of the radial coordinate $r$
for the sine-Gordon case.
Second column: reduced action density $\sigma$ as a function of $r$.
Third column: $2\pi\int_0^R \sigma(r)\,rdr$ as a function of $R$.
Fourth column: azimuthal component $J_\theta$ of the Abelian current $\mathbf{J}$ for
the sine-Gordon case.
}
\label{figsine}
\end{center}
\end{figure}

Equation~(\ref{EDsine}) may be investigated along the same lines as in section~\ref{subsection:sinh}.
Once again expanding $\alpha(r)$ for small $r$ leads to an equation similar to \eq{alphaas}, namely
\be
\label{alphaas.sine}
\alpha(r) \xrightarrow[r\to0]{}
a + \frac{\kappa^2}{8} \sin(2a) \, r^2 + \frac{\kappa^4}{256} \sin(4a) \, r^4 + \mathcal{O}(r^6),
\ee
which can be used to generate initial conditions for $\alpha(r)$ and $\alpha'(r)$ at
$r\approx0$ (we note that exact asymptotic expressions, similar to
equations (\ref{exact.sinh}) and (\ref{exact.sinh.parameters}), are also available
for this case$^{\ref{fn1}}$~\cite{novo1,novo2,novo3,novo4,book}).
This, in turn, may be used to numerically integrate \eq{EDsine}.
Proceeding similarly to the previous case, we start with $\alpha(10^{-3})$ and $\alpha'(10^{-3})$
defined by \eq{alphaas.sine}, with $\kappa=1$ and $a=\frac{3\pi}{4}$.
This yields the plots of Fig.~\ref{figsine}.
A noticeable difference with the sinh-Gordon case is that the solutions are
not localized in any sense here, as shown by the comparison between Figs.~\ref{figsinh} and \ref{figsine}.
This is further illustrated in Fig.~\ref{figFz3d}, where we compare the Abelian field
strength $F_{12}$ obtained for the sinh-Gordon and sine-Gordon cases.

\section{Smooth solutions in $T^2$ for $G=SO(2,1)$}

Finally, we remark that the elliptic sinh-Gordon equation \eq{ex.sinh.b}
admits smooth doubly-periodic solutions on the 2-torus $T^2$. In \cite{tcl-prl},
the authors show that \eq{ex.sinh.b} appears in problems related to fluid and plasma
physics and, in~\cite{tcl}, the same authors construct solutions in closed form in
terms of Riemann theta functions. This kind of solutions to \eq{ex.sinh.b},
given in terms of $\theta$-functions, are extensively analyzed in \cite{B}.
Evidently, such solutions may in turn be regarded as smooth solutions of
the $SO(2,1)$ Hitchin's equations on the 2-torus $T^2=\RR^2/\ZZ^2$.

To describe the solutions, we must consider a smooth hyperelliptic curve
$C\subset\CC^2$ of genus $g\ge1$, known as the spectral curve, which is given by:
$$ C = \{ (u,w)\in\CC^2 ~|~
u^2 = w \cdot \Pi_{k=1}^g(w-\lambda_k)(w-\overline{\lambda_k}^{-1}) \} , $$
where the parameters $\lambda_1,\dots,\lambda_g$ are all distinct and
satisfy $|\lambda_k|\ne1$. Now let $\theta:\CC^g\to\CC$ be the Riemann
$\theta$-function for the curve $C$ \cite{GH}.
As it is shown by Bobenko \cite{B}, every doubly-periodic, non-singular, real-valued
solution of the elliptic sinh-Gordon equation (\ref{ex.sinh.b}) is of the form
\be\label{dp-soln}
\alpha(z,\overline{z}) = 2\log \left(
\frac{\theta(-i(Uz+V\overline{z})/2+D)}{\theta(-i(Uz+V\overline{z})/2+D+\Delta)} \right)
\ee
where $D$ is an arbitrary imaginary vector in $\CC^g$, $\Delta=i\pi(1,\dots,1)\in\CC^g$,
and $U,V\in\CC^g$ are the period vectors, which are uniquely determined by the curve $C$.

For a fixed value of $g\ge1$, notice that the set of doubly-periodic, non-singular,
real-valued solutions of (\ref{ex.sinh.b}) depends on $3g$ real parameters: the $g$
complex parameters $\lambda_1,\dots,\lambda_g$ (which determine the spectral
curve $C$ and therefore determines the $\theta$-function and the vectors $U,V$)
plus $g$ real parameters determining the imaginary vector $D\in\CC^g$.
We therefore conclude that the moduli space of smooth solutions the $SO(2,1)$ Hitchin's
equations on $T^2$ posesses an infinite number of connected components which are parametrized
by a positive integer $g\ge1$ (the genus of the spectral curve), and that each connected
component has dimension at least $3g$.

\section{Conclusion}

In this paper, we have discussed smooth real solutions of the $SO(2,1)$ Hitchin's
equations on $\RR^2$, $S^2$ and $T^2$, even though there are no such solutions for
the $SU(2)$ Hitchin's equations. This was done by using an ansatz that reduced the
$SO(2,1)$ Hitchin's equations to three classical integrable equations.

We believe that our calculations provide a strong motivation for further study of Yang-Mills
theory for noncompact real Lie groups. We provided evidence that some of the good
properties of Yang-Mills theory for compact real Lie groups, like the possible existence
of finite dimensional moduli spaces and the link with integrability, are preserved when
noncompact real Lie groups are considered. We propose that a general Yang-Mills theory
for complex Lie groups and its noncompact real forms ought to be developed.

\acknowledgments
We thank an anonymous referee for calling our attention to several
exact results concerning the global asymptotic analysis of the
third Painlev\'e equation.
We are grateful to G. Etesi for several discussions, and to
L. Bonora and F. Williams for drawing our attention to~\cite{bonora} and \cite{tcl}.
RAM acknowledges FAPESP for financial support (grant 04/06721-2).
MJ is partially supported by the CNPq grant number
300991/2004-5 and during the preparation of this paper also
received funding from the FAEPEX grants number 1433/04
and 1652/04.

\end{document}